\documentclass[conference]{IEEEtran}

\usepackage{cite}
\usepackage{amsmath,amssymb,amsfonts}
\usepackage{algorithmic}
\usepackage{graphicx}
\usepackage{textcomp}
\usepackage{xcolor}
\usepackage{xspace}
\usepackage{booktabs}
\usepackage{multirow}

\newif\ifshowcomments
\showcommentstrue % turn ON - comment this line to turn them off

\newcommand{\tool}{\texttt{Pomona}\xspace}
\newcommand{\scanFlow}{\texttt{Scanning}\xspace}
\newcommand{\repFlow}{\texttt{Repair}\xspace}
\newcommand{\textQuote}[1]{``\textit{#1}''\xspace}

\begin{document}

\title{Pomona: Continuous Code Quality Improvement via Small, Agentic Pull Requests at Bloomberg}

\author{\IEEEauthorblockN{David Williams}
\IEEEauthorblockA{\textit{University College London}\\
London, United Kingdom \\
david.williams.22@ucl.ac.uk}
\and
\IEEEauthorblockN{Angelos Evripiotis}
\IEEEauthorblockA{\textit{Bloomberg}\\
London, United Kingdom \\
jevripiotis@bloomberg.net}
\and
\IEEEauthorblockN{Serkan Kirbas}
\IEEEauthorblockA{\textit{Bloomberg}\\
London, United Kingdom \\
skirbas@bloomberg.net}
\and
\IEEEauthorblockN{Sergey Magidovich}
\IEEEauthorblockA{\textit{Bloomberg}\\
London, United Kingdom \\
smagidovich@bloomberg.net}
\and
\IEEEauthorblockN{Harry Morgan}
\IEEEauthorblockA{\textit{Bloomberg}\\
London, United Kingdom \\
hmorgan39@bloomberg.net}
\and
\IEEEauthorblockN{Peter Wainwright}
\IEEEauthorblockA{\textit{Bloomberg}\\
London, United Kingdom \\
pwainwright@bloomberg.net}
\and
\IEEEauthorblockN{Federica Sarro}
\IEEEauthorblockA{\textit{University College London}\\
London, United Kingdom \\
f.sarro@ucl.ac.uk}
}

\maketitle

\begin{abstract}
In this industrial experience paper, we present \tool, a lightweight agentic tool that utilises agent skills for continuous code quality improvement. Inspired by the Kaizen\texttrademark{} philosophy, \tool automates a cycle of discovery and incremental repair: a \scanFlow skill identifies tasks and prioritises them in a backlog, while a \repFlow skill generates small, easily reviewable pull requests (PRs). This design enables frequent, low-risk improvements while maintaining engineer trust and reducing technical debt. We evaluated \tool at Bloomberg through a three-month team deployment and a questionnaire distributed to senior engineers. The results are promising: 32/39 PRs (82.1\%) were merged with a median time-to-close of just over two hours. Moreover, 10/12 surveyed engineers expressed a desire to adopt \tool, praising its small diff sizes and focus on code quality. Since our evaluation, another team has adopted \tool. We conclude with actionable insights for deploying agents in industry.
\end{abstract}

\begin{IEEEkeywords}
Software Quality, Technical Debt, AI4SE, Agent
\end{IEEEkeywords}

\section{Introduction}
As Large Language Model (LLM) integrations and agentic solutions~\cite{yang_swe-agent_2024, ruan_specrover_2025, wang_openhands_2025} mature in the domain of Software Engineering (SE), organisations are increasingly looking to implement reliable techniques to improve the software development process for their employees. Although some studies on the impact of LLM-based SE tools on developer productivity have reported significant increases in development speed~\cite{peng_impact_2023}, others have observed that developers using LLMs tend to perform their regular tasks more slowly than without (while believing they are working faster)~\cite{becker_measuring_2025}. Notably, He et al.~\cite{he2026speedcostqualitycursor} found that agent adoption in open source repositories increases development velocity only temporarily and comes at the cost of longer-term increases in code complexity and quality issues.

\begin{figure*}[t]
\centerline{\includegraphics[width=0.95\linewidth]{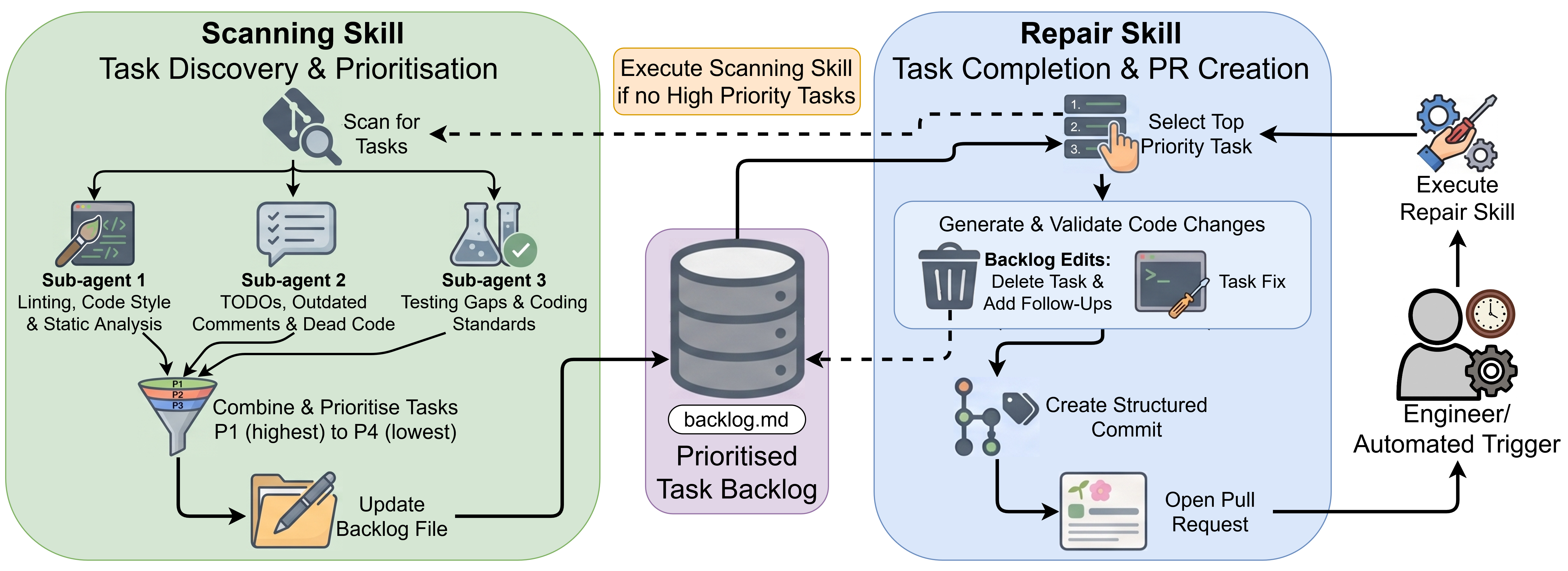}}
\caption{Overview of \tool. Responsibilities are divided into two agent skills, each of which interacts with a structured task backlog. The \scanFlow skill (left) discovers tasks and populates the backlog, while the \repFlow skill selects the highest-priority task and generates a small, easily reviewable pull request.}
\label{fig:sys_overview}
\end{figure*}

To reconcile this tension between short-term speed and long-term maintainability, we present a novel, Kaizen\texttrademark-inspired\footnote{Kaizen\texttrademark{} is a Japanese word and business methodology promoting continuous improvement through incremental, consistent changes~\cite{imai1986kaizen}.} agentic solution, dubbed \tool, that utilises agent skills to improve code quality continuously through small, automated changes. Agent skills are user-defined instructions that enable autonomous agents to handle specialised, repeatable workflows~\cite{copilot-skills, claude-skills}. \tool comprises two such skills: one for identifying code quality improvement tasks and another for addressing them. It builds and maintains a structured, prioritised backlog of these tasks, which it gathers from multiple sources defined by the adopting user, e.g., static analysers, inline technical debt markers, test coverage gaps, and deviations from project-specific coding standards. Picking the highest-priority item from the backlog each time, \tool generates tiny pull requests (PRs) targeting approximately 10 lines of diff, then updates the backlog and repeats the process. \tool handles the needs and challenges of efficiently adopting AI by being \textcircled{1} low friction, in that the task discovery and patch generation processes are autonomous background workflows that do not require engineers to adapt their development activities; \textcircled{2} low stakes, adopting a human-in-the-loop review paradigm so that \tool's contributions cannot affect the codebase without receiving engineer approval (with that process made especially simple for the engineer given the small size of \tool's PRs); and \textcircled{3} focused on improving code quality, ensuring that changes introduced by \tool are highly likely to benefit the codebase's long-term maintainability.

\tool is already in use at Bloomberg, with positive early impressions.
To evaluate its potential value in practice, we conducted a mixed-methods study, beginning with a manual audit of 39 PRs generated by \tool for the first adopting team over the course of a three-month deployment. We observed both high acceptance and efficiency: 32/39 (82.1\%) of the generated PRs were successfully merged, with 25 of those accepted requiring no human interaction beyond the mandatory review, all within a median time-to-close of 2 hours and 14 minutes. We complement these findings with insights from a questionnaire distributed to 12 senior engineers not yet using \tool. The results are promising, with 10/12 participants expressing a desire to try the tool, particularly valuing \tool's small diff sizes (10/12) and focus on improving code quality (10/12). At the time of writing, a second engineering team has adopted \tool, and plans are in place to distribute it more widely within Bloomberg. In the remainder of this paper, we provide an in-depth description of \tool's implementation and offer practical insights and strategies for deploying agentic AI4SE in a large-scale industrial environment.

\section{\tool Overview}
Figure~\ref{fig:sys_overview} illustrates an overview of \tool, which comprises two agent skills: one for identifying code quality improvement tasks in the target repository and compiling them in a structured, prioritised backlog (the \scanFlow skill, \S\ref{sec:scanning_workflow}); and another for selecting the highest-priority task in the backlog and generating a new PR that addresses it for human review (the \repFlow skill, \S\ref{sec:repair_workflow}). Leveraging the agent skill medium makes \tool's implementation simple: the \repFlow skill, the \scanFlow skill, and the structured task backlog are each implemented as individual Markdown files that can be integrated with any current agentic product. This format also makes \tool programming-language-agnostic (although our specific implementation, evaluated in \S\ref{sec:eval_audit}, targets Python).
The following sections describe each component and their cyclical combination to achieve a continuous code-quality improvement loop.

\subsection{Prioritised Task Backlog} \label{sec:tool_backlog}
In \tool, a structured backlog of code quality improvement tasks lies at the centre of the \scanFlow and \repFlow skills. The \scanFlow skill fills the backlog with prioritised improvement tasks, while the \repFlow skill consumes these tasks to generate new PRs with improvements for human review. An excerpt from the backlog is shown in Figure~\ref{fig:backlog}.

Backlog improvement tasks are categorised in priority groups P1--P4 defined in the ``benefit $\times$ ease-of-review'' matrix in Table~\ref{tab:priority-matrix}. By altering the specification of the \scanFlow skill, engineers adopting \tool can define what constitutes \textit{high benefit} in their development context. A good rule of thumb is that bug-catching improvements should take priority over cosmetic ones. For example, a rule that prevents mutable default parameters is more valuable than one that reorders imports. For the specific implementation of \tool we evaluated in \S\ref{sec:eval_audit}, we included the following in our \scanFlow skill specification: ``\textit{high-benefit} changes catch real bugs or prevent future bugs (mutable defaults, loop variable capture, missing exception chains); reduce maintenance burden (removing dead code, fixing misleading comments); or improve developer experience (better error messages, clearer code)''.

Following a similar process to the one described above, adopting engineers can also define what counts as \textit{easy to review}. For our implementation, we included the following rule in our \scanFlow skill: ``A change is easy to review if it is fully automated (auto-fix, formatting), mechanical and repetitive (reviewer can scan quickly), small in scope (single file or rule) or does not introduce behavioural changes''.

Our initial design for \tool stores the task backlog in a Markdown file in the repository root. While this design provides some benefits (i.e., the backlog is centralised and versioned alongside the codebase), we note that the backlog could also be implemented using project management software such as Jira via Model Context Protocol (MCP) integrations.

\subsection{\scanFlow: Identifying \repFlow Tasks} \label{sec:scanning_workflow}
The \scanFlow skill automatically discovers and prioritises new code quality improvement tasks. It is executed as part of the \repFlow skill (\S\ref{sec:repair_workflow}), specifically when the backlog runs low on high-priority tasks (the P1 and P2 categories are empty). To refill the backlog, the \scanFlow skill initialises multiple sub-agents in parallel, each exploring a different discovery area. As described below, the quantity and purpose of sub-agents should be defined by the adopting user based on their needs. Once all sub-agents complete their searches, their outputs are combined into prioritised backlog tasks.

In our specific implementation of \tool, three unique sub-agents are defined. Sub-agent 1 performs rule expansions for \texttt{ruff} (a Python linter and code formatter~\cite{ruff}) and strict type-checking expansion for \texttt{mypy} (a static type checker for Python~\cite{mypy}). Sub-agent 2 searches the codebase for TODOs expressed as ``\texttt{TODO|FIXME|HACK|XXX}'', and classifies each finding according to the following guide: self-documenting removals (where the code is already done) are assigned as P1, clear fixes with obvious solutions are P1/P2, fixes requiring further domain investigation are P3, and aspirational comments are assigned as P4. Sub-agent 2 also searches for comments that contradict their surrounding code, as well as identifies dead code by searching for unused imports, commented-out code blocks ($>$3 consecutive commented lines), and unreachable code after early returns. Finally, sub-agent 3 identifies test coverage gaps by comparing source modules with unit test modules to reveal untested modules. It prioritises pure-logic modules without data dependencies (as these are easiest to test) and modules with complex branching logic (as these provide the highest value). In addition, sub-agent 3 checks compliance with the repository's defined coding standards (usually found in agent configuration files) and looks for long functions (i.e., longer than 50 lines) that can be decomposed into smaller parts.

Once all agents complete their searches, the \scanFlow skill concludes by aggregating their findings. This aggregation is done in two steps: First, the findings are listed and concatenated, and duplicate findings are removed. Second, each unique finding is assigned to a priority category following the priority matrix shown in Table~\ref{tab:priority-matrix} and rules described in \S\ref{sec:tool_backlog}.

\begin{figure}[tb]
\centerline{\includegraphics[width=0.8\linewidth]{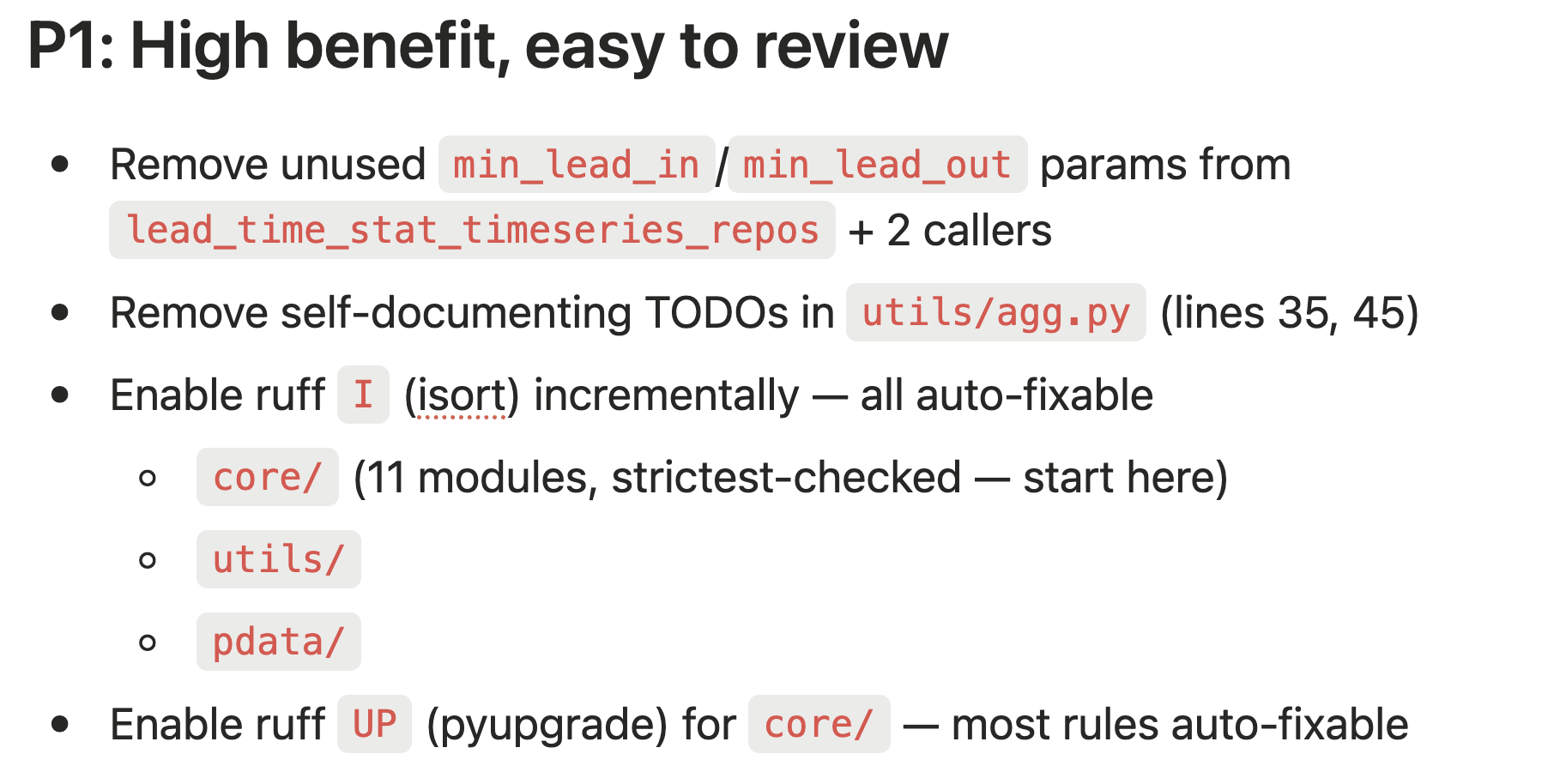}}
\caption{Excerpt of \tool's backlog for the adopting Bloomberg engineering team featured in \S\ref{sec:eval_audit}.}
\label{fig:backlog}
\end{figure}

Now with priority levels assigned, tasks are converted to follow the backlog item format (shown in Figure~\ref{fig:backlog}) and added to their corresponding priority category. At this point, these tasks are ready to be handled by the \repFlow skill, which is described in the following section.

\begin{table}[tb]
\caption{Priority Matrix for Task Management}
\begin{center}
    \resizebox{0.85\columnwidth}{!}{
    \begin{tabular}{|l|l|l|}
        \hline
         & \textbf{Easy to review} & \textbf{Hard to review} \\ \hline
        \textbf{High benefit} & \textbf{P1} --- Do first & \textbf{P3} --- Worth doing, plan the split \\ \hline
        \textbf{Low benefit} & \textbf{P2} --- Quick wins & \textbf{P4} --- Parking lot (do not pick) \\ \hline
    \end{tabular}
    }
\label{tab:priority-matrix}
\end{center}
\end{table}

\subsection{\repFlow: Solving Tasks Identified by \scanFlow} \label{sec:repair_workflow}

The \repFlow skill is \tool's point of entry, implemented as an agent skill that can either be executed by the user or configured to trigger periodically (e.g., once per day).

The first step of the \repFlow skill instructs the agent to access the task backlog at the root of the target repository. If the backlog file does not exist or the high-priority (P1 and P2) task categories are empty, \textit{the \scanFlow skill (\S\ref{sec:scanning_workflow}) is executed first}, after which the agent returns to complete the remaining steps in the \repFlow skill.

When there are tasks in the backlog, the next step is to \textit{select the first task from the highest-priority category}. If an item has sub-items (as in Figure~\ref{fig:backlog}), the instructions specify to select the first sub-item.

The next step is to \textit{implement the fix}. For this, a few rules are defined. First, any changes should be validated with the project's testing and linting commands (as specified by the adopting team). Second, changes must be small: since the goal is to create tiny, easily reviewable PRs, the agent should aim for roughly 10 lines of diff. If a change exceeds that, the agent is instructed to split it using strategies provided by the adopting team. For example, in our implementation, one strategy is to enable linting rules/add tests for one directory at a time, and then add follow-up items for other directories to the backlog.

Once the code changes are made and validated, the agent \textit{updates the backlog} by deleting the completed task and adding follow-up tasks (if any). Since the backlog is stored in the repository, there is no need to maintain a separate log of completed tasks, as the project's \texttt{git} history tracks changes made to the backlog over time.

With the code and backlog changes complete, the agent can now \textit{commit the changes}, including a clear motivation for the change in the commit body and links to relevant resources to help human reviewers better understand these automated changes when reviewing the project's history down the line.

The final step of the \repFlow skill \textit{asks the user whether they want to generate a PR}. If so, the agent creates one (our implementation uses Bloomberg's internal MCP integrations to achieve this) following the repository's standard PR template. This choice is optional depending on the control paradigm: if the skill is autonomously triggered periodically (e.g., once per day) rather than initialised by an engineer, this user choice is skipped, and a PR is created as soon as the previous steps are complete. The PR title is generated as a one-sentence summary of what was improved and why it matters, prefixed with specific emojis to clearly indicate that the PR is AI-generated. For the PR description, the agent is instructed to lead with the concrete outcomes for engineers. It only includes additional sections if they add value beyond the title. For example, it would include the ``Improvements'' section only if the title alone does not sufficiently explain the changes, and the ``Testing \& Linting'' section only if special testing was done beyond the project's standard testing commands.

\section{Early Insights and Evaluation} \label{sec:evaluation}
We conducted a mixed-methods evaluation comprising an analysis of PRs generated by \tool for one team over a three-month adoption period and a questionnaire distributed to non-\tool users gauging their interest in its concept.

\subsection{Analysing \tool Pull Requests} \label{sec:eval_audit}
We assessed \tool's early-stage impact by analysing PRs generated by \tool for the first adopting team over a three-month period. During this time, \tool generated 39 PRs, all of which were closed by the team (either merged or rejected). We manually extracted standardised metadata for each PR, including the resolved task's priority level (P1--P4) and nature (i.e. source). To measure the complexity of each PR, we recorded the diff size (files and lines of code) and whether the changes led to additional subtasks being added to the backlog. Finally, we observed the outcomes and review dynamics for each PR, including whether the PR was merged, the time between creation and closing, the number of reviews and non-review comments, and whether any follow-up commits were required before merging. For rejected PRs, we noted any explicit reasoning provided.

\subsubsection{Results}

Our evaluation period spans approximately three months from mid-March to mid-June 2026. The team preferred to retain full control over the volume of AI-generated PRs in their repository, so a team member manually executed the \repFlow skill for each new PR (allowing the \scanFlow skill to determine the type and order of addressed tasks). The team aimed for a goal of ``one \tool PR per day'', which they largely achieved, producing 39 PRs across 55 workdays (accounting for public and team holidays). Across these PRs, we observed a strong overall acceptance rate of 82.1\% (32/39). 

Regarding backlog priority levels, 31 PRs generated by \tool were P1, and the remaining 8 were P2. Although tasks existed in the other priority-level groups, there were sufficient P1- and P2-level tasks such that the other priority groups were not reached within the three-month evaluation period. The majority (26/39) of the \tool PRs addressed \texttt{ruff} rule violations, and other tasks included adding unit tests (5 PRs), removing dead code (3 PRs), and other repository-specific tasks (5 PRs). Notably, 6 PRs were directly requested by a team member, who manually added the corresponding backlog entries in different PRs. This example highlights the flexibility offered by \tool's accessible backlog in the repository root.

Taking a closer look at the seven rejected PRs reveals that four were due to ``race conditions'', in which \tool was executed twice before a human review, resulting in duplicate PRs. Each duplicate was rejected, while the original was accepted. Taking this oversight into account, we revised \tool to skip tasks that are already addressed by currently open PRs. Of the remaining three rejections, two were due to changes being made with undesirable formatting, and the final one was closed without comment from the team.

Of the 32 accepted PRs, 28 were accepted as-is without requiring additional commits from the human reviewers, and 25 were accepted with no comments or interactions beyond the mandatory human review before merging. This largely positive outcome suggests that the scope of tasks tackled by \tool was manageable enough to typically warrant only a brief review. Reflecting this low complexity, the median time to close (either merge or reject) for \tool PRs was only 2 hours and 14 minutes. PR turnaround was consistently swift: 25 of the 39 PRs (64.1\%) were resolved within 4 hours, increasing to 29 (74.4\%) within the same day and 33 (84.6\%) by the next business day. While we do observe some outliers at both ends, ranging from 1 minute when duplicate PRs were closed to a maximum of four business days during periods where the team's reviewing capacity was particularly low, the overall distribution of time-to-close measurements supports our claim that \tool is low-friction and integrates into existing workflows without significant cognitive overhead.

\tool PRs exhibited a substantial range in size. All PR size metrics (files/lines changed) exclude backlog changes to focus solely on codebase impact. The median number of files changed was 4 (mean 6.5), with a minimum of 1 and a maximum of 35. One potential cause for concern was the number of lines changed, calculated as the sum of added and deleted lines. The median was 37, the mean was 57.7, and values ranged from 5 to 168. While our original implementation of the \repFlow skill explicitly specified 10-line diffs, the agent sometimes failed to adhere to this limit when handling repetitive refactoring tasks. Despite this deviation, the team merged more than half of the nine PRs over 100 lines within the same business day, indicating that even larger-scale PRs can be manageable when the targeted improvements are intuitive. Here, larger PRs typically applied a single \texttt{ruff} rule to an entire file or subdirectory at once. Based on these empirical insights, the adopting team has since increased \tool's PR size limit to 100 lines and enforced it with a deterministic check that automatically rejects PRs exceeding that limit. Noting the limitations of size as a metric, we highlight the need for better proxies to estimate PR review effort as an important direction for future work.

Finally, six of the accepted \tool PRs added new tasks to the backlog, including one that triggered the \scanFlow skill, which added 8 new tasks. The other five PRs added sub-tasks directly related to the task at hand to ensure the scope of changes did not become too large to review easily.

\subsection{Non-User Questionnaire} 
\label{sec:eval_survey}
To gauge interest in \tool's concept, we created a questionnaire to distribute among engineers. We specifically targeted engineers in Bloomberg's Developer Experience (DevX) teams, given their rich understanding of the organisation's development tooling. The questionnaire contained 12 questions, the first five of which covered participant demographics, including job title, years of experience, and familiarity with AI4SE tools. Next, the questionnaire presented the participant with Figure~\ref{fig:sys_overview} alongside a high-level description of \tool, followed by seven questions concerning whether the tool seems useful (and if not, why), how many pull requests the participant would be willing to review per week if they adopted the tool, which features seem most valuable, what types of tasks they would want \tool to tackle, and concluding with an opportunity to provide any additional feedback to refine the concept. We provide the questionnaire as part of our supplementary materials~\cite{repo}.

\subsubsection{Results}
In total, 12 participants completed the questionnaire. This cohort comprised two Team Leads, nine Senior Software Engineers and one software engineer, of whom seven had more than 10 years of experience, four had 4--10 years of experience, and one had 1--4 years of experience. In terms of experience with AI-assisted software development (i.e., in-IDE tools), nine stated they use these tools daily, one weekly, another monthly, and one never. Similarly, six participants stated they use agentic tools (e.g., automated PR generation and review) daily, two weekly, three monthly, and one never.

When presented with \tool's concept, the majority (9/12) of participants felt the tool would be useful, and more (10/12) stated they wanted to try \tool in future. If they adopted \tool, most participants (8/12) would be willing to review 2--3 PRs per week, which is only a little less than the number observed in the adopting team described in \S\ref{sec:eval_audit} (3--4 PRs per week). The remaining responses were split: one participant was willing to review more than 5 PRs per week, another would review 4--5 per week, and the remaining two preferred 1 or fewer per week. When asked about the most appealing features to them among those described in the concept, small diffs and focus on code quality improvement were both praised by 10/12 participants, while 7/12 placed importance on the automated identification and completion of tasks. These responses suggest most participants appreciate \tool's format but require high confidence and control over the automated task selection and prioritisation process.

In terms of tasks the participants wish \tool would tackle, six requested handling TODOs/outdated comments, six mentioned static analysis/linting issues, three mentioned simple bug fixing, and another three mentioned dependency management, seemingly because they trust agents to handle issues of that scope autonomously. Others noted documentation, testing (which would be the \textQuote{one exception to the size [limit] of the PR}), performance fixes, and removing dead code.

A few participants also highlighted potential drawbacks and areas for future improvement. They highlighted the risk of review overload, fearing that \tool might simply \textQuote{create more open PRs for actual humans to handle}, or generate \textQuote{gratuitous churn} by addressing tasks that, while technically valid, are \textQuote{not relevant enough to care about}. Another participant noted that \textQuote{sometimes code smells require large-scale refactoring which seems to be out of scope [for \texttt{{Pomona}}]}. However, the fact that 10/12 participants still expressed a desire to try the tool suggests that, despite their smaller scope, the aforementioned code quality improvement tasks are primarily neglected due to a lack of engineering bandwidth rather than a lack of perceived value. In other responses, trust and agency emerged as key topics of interest. Several participants were wary of \textQuote{trust[ing] the AI judgement to discover the right issues to fix} and expressed a preference for a user-controlled paradigm (i.e., a fully user-triggered \repFlow skill, as described in \S\ref{sec:eval_audit}) over an autonomous triggering system, with this also extending to wanting \textQuote{control over what gets prioritised first}. Finally, one respondent requested the ability for \tool to iterate upon its own PRs based on reviewer feedback. We consider agentic PR refinement a promising area for future work.

\section{Discussion}
Based on our experience in deploying and evaluating \tool at Bloomberg, we discuss actionable insights into industrial adoption of agentic AI and outline future work for both researchers and practitioners.

\noindent \textbf{Small Changes Facilitate Adoption.} 
The early success of \tool was not in solving the most ``difficult'' coding problems, but in tackling high-priority, low-complexity tasks with high accuracy and minimal friction.
The high PR acceptance rate in our first industrial deployment suggests that current agents can confidently tackle ``low-hanging fruit'' in software repositories (e.g., linting violations, dead code). Small, intuitive changes reduce both reviewer scepticism and cognitive load. In fact, despite \tool's strong performance, the adopting team intentionally limited its output to at most 1 PR per day, indicating that human review bandwidth is a far greater bottleneck in AI-driven development than code-generation speed. Moreover, 10/12 survey participants cited small diffs as \tool's most appealing feature, highlighting the importance of code reviewability in accepting AI contributions.

\noindent \textbf{AI for Technical Debt Management.} 
Our findings suggest that engineers want AI to handle "neglected" tasks (TODOs, outdated comments, dead code, and linting) because while these tasks are important, engineers struggle to find time to address them manually. \tool demonstrates that agents can easily be configured to target TODOs, outdated comments, and dependency updates. These are the specific areas where the engineers in our study strongly valued automation. The fact that several \tool PRs generated additional tasks also shows that AI can act as a catalyst for ongoing technical debt management rather than just a one-off tool.

\noindent \textbf{Prioritisation is as Important as Detection.} While static analysis tools can identify many potential tasks, selecting the right improvements to apply is crucial. Our ``benefit $\times$ ease-of-review'' prioritisation strategy (Table~\ref{tab:priority-matrix}, \S\ref{sec:tool_backlog}) enables agents to focus on high-benefit, low-friction changes, ensuring that automated changes deliver tangible value. Given the bottlenecks in human code-reviewing bandwidth, an avenue for future research could be automated task-relevance filtering, developing techniques that can distinguish ``relevant'' or ``important'' code changes from merely ``valid'' ones.

\noindent\textbf{Agentic Task Decomposition}. Another direction for future work to facilitate industry AI adoption would be to investigate how agents can reliably identify and break down large tasks into small, easily reviewable ones without human intervention.

\noindent \textbf{Expert Scepticism and Human-in-the-Loop.}
Through \tool's PR-based format, engineers retain full control over whether changes enter their codebases. These PRs clearly communicate the purpose and impact of each modification, following consistent templates that prioritise clarity and benefit for the engineer. This transparency is critical for building trust in automated changes. Our survey reveals that experienced engineers (10+ years' experience) remain wary of ``agentic judgment'' in prioritising tasks. For this reason, agentic solutions should continue to promote a user-centric~\cite{williams-2024-bbapr} paradigm such as that of \tool, with its user-initiated workflow, accessible backlog, and review-dependent impact.

\noindent\textbf{Quantifying Cognitive Load.} Currently, it is difficult to identify the threshold at which agentic PRs transition from helpful suggestions to ``gratuitous churn''. Future research should shift from binary ``merge/reject'' rates toward multi-dimensional metrics (incorporating subjective engineer perceptions) that capture the cognitive load of human-AI collaboration. This effort requires devising proxies to measure the context-switching cost of reviewing AI and human code, as well as the mental effort involved in verifying automated changes.

\section{Related Work}
Preliminary research on agent skills in different domains has explored their effectiveness in improving task performance, including SE tasks~\cite{li2026skillsbench, han2026sweskillsbench}, and how to organise~\cite{liang2026skillnet}, secure~\cite{liu2026wild, liu2026maliciouswild}, and evolve skills at scale~\cite{alzubi2026evoskillautomatedskilldiscovery, SkillMOO26}. The results generally show that curated skills can improve performance~\cite{li2026skillsbench}, especially in specific scenarios~\cite{han2026sweskillsbench}, although the gains in SE tasks remain context-dependent~\cite{li2026skillsbench} and their overall impact is still unclear~\cite{han2026sweskillsbench}. These results show the challenges of designing effective, reusable procedural knowledge for SE tasks, and the need for governance and validation mechanisms when deploying such skills in practice \cite{liu2026wild, liu2026maliciouswild}.

In contrast to previous work, which primarily focuses on evaluating, generating, or securing agent skills, our work investigates how to devise and integrate agent skills in a real-world industrial setting based on user needs. The tool we present uses skills to enable low-friction, continuous, incremental code-quality improvement, and we highlight its value to users through an early adoption study. In addition, we report practical lessons from its deployment.

\section{Conclusion}
This work introduces \tool, an agentic tool that leverages agent skills for continuous code quality improvement through automated, easily reviewable changes. Our early evaluation, including a three-month team deployment and a questionnaire distributed to senior practitioners, demonstrates \tool's potential to help manage technical debt with minimal impact on development velocity, and to serve as a low-stakes, high-benefit approach to agentic AI adoption in SE.

\section*{Data Availability Statement}
As \tool has been developed and evaluated within Bloomberg's internal systems and repositories, we cannot release the data and source code due to non-disclosure agreement policies. We also cannot reveal which AI services or models were used. However, \tool's implementation consists of three Markdown files (the \repFlow skill, the \scanFlow skill, and the backlog) and can be combined with any language model. In this work, we have detailed how to implement such a tool in practice, enabling academics and practitioners alike to adopt it in future. Moreover, we make the questionnaire we distributed for user evaluation available~\cite{repo}.

\bibliographystyle{IEEEtran}
\bibliography{IEEEabrv,references}

\end{document}